\documentclass[aps,prl,twocolumn,showpacs,groupedaddress,floatfix]{revtex4}  % for review and submission 
\usepackage{graphicx}  % needed for figures
\usepackage{dcolumn}   % needed for some tables
\usepackage{bm}        % for math
\usepackage{amssymb}   % for math
\usepackage{xspace}

\begin{document}
\newcommand{\dzero} {D\O\xspace} \newcommand{\ttbar}
{\ensuremath{t\bar{t}}\xspace} \newcommand{\ppbar}
{\ensuremath{p\bar{p}}\xspace} \newcommand{\thetad}
{\ensuremath{\theta^{*}}\xspace}
\newcommand{\costheta}{\ensuremath{\cos\thetad}\xspace}
\newcommand{\ljets}{lepton+jets\xspace}
\newcommand{\mujets}{\ensuremath{\mu+jets}\xspace}
\newcommand{\ejets}{\ensuremath{e+jets}\xspace}
\newcommand{\fplus}{\ensuremath{f_{+}}\xspace}
\newcommand{\fminus}{\ensuremath{f_{-}}\xspace}
\newcommand{\fzero}{\ensuremath{f_{0}}\xspace}
\newcommand{\wjets}{\ensuremath{W+}jets\xspace} \newcommand{\met}
{\mbox{$\not\!\!E_T$}\xspace} \newcommand{\tld}{\ensuremath{{\cal
      D}}\xspace} % symbol for the topological likelihood discriminant
\newcommand{\tWb}{\ensuremath{t \rightarrow
    Wb}\xspace} % symbol for the topological likelihood discriminant
\newcommand{\SM}{standard model\xspace} \newcommand{\MC}{Monte
  Carlo\xspace} \newcommand{\CL}{C.L.\xspace}
\newcommand{\deteta}{\ensuremath{\eta_{\mathrm{det}}}}
\newcommand{\alpgen}{{\sc{alpgen}}\xspace}
\newcommand{\pythia}{{\sc{pythia}}\xspace}
% words that need special hyphenation
\hyphenation{an-aly-sis}

% the following information is for internal review, please remove them
% for submission \leftline{Version 1.2 as of \today} \leftline{Primary
%   authors: Bryan Gmyrek, Christian Schmitt} \rightline{Comment to
%   {\tt d0-run2eb-022@fnal.gov}} \rightline{by xxx, 2005}

% the following line is for submission 
\hspace{5.2in}
\mbox{Fermilab-Pub-05/187-E}

\title{Search for right-handed W bosons in top quark decay}
% LIST_OF_AUTHORS_R2.TEX                 5/10/05            
%
\author{                                                                      
%% names begin here                                                           
V.M.~Abazov,$^{35}$                                                           
B.~Abbott,$^{72}$                                                             
M.~Abolins,$^{63}$                                                            
B.S.~Acharya,$^{29}$                                                          
M.~Adams,$^{50}$                                                              
T.~Adams,$^{48}$                                                              
M.~Agelou,$^{18}$                                                             
J.-L.~Agram,$^{19}$                                                           
S.H.~Ahn,$^{31}$                                                              
M.~Ahsan,$^{57}$
% ...
\\                                                                            
\vskip 0.30cm                                                                 
\centerline{(D\O\ Collaboration)}                                             
\vskip 0.30cm                                                                 
}                                                                             
\affiliation{                                                                 
\centerline{$^{1}$Universidad de Buenos Aires, Buenos Aires, Argentina}       
\centerline{$^{2}$LAFEX, Centro Brasileiro de Pesquisas F{\'\i}sicas,         
                  Rio de Janeiro, Brazil}                                     
\centerline{$^{3}$Universidade do Estado do Rio de Janeiro,                   
                  Rio de Janeiro, Brazil}                                     
\centerline{$^{4}$Instituto de F\'{\i}sica Te\'orica, Universidade            
                  Estadual Paulista, S\~ao Paulo, Brazil}                     
\centerline{$^{5}$University of Alberta, Edmonton, Alberta, Canada,           
               Simon Fraser University, Burnaby, British Columbia, Canada,}   
\centerline{York University, Toronto, Ontario, Canada, and                    
         McGill University, Montreal, Quebec, Canada}                         
\centerline{$^{6}$Institute of High Energy Physics, Beijing,                  
                  People's Republic of China}                                 
\centerline{$^{7}$University of Science and Technology of China, Hefei,       
                  People's Republic of China}                                 
\centerline{$^{8}$Universidad de los Andes, Bogot\'{a}, Colombia}             
\centerline{$^{9}$Center for Particle Physics, Charles University,            
                  Prague, Czech Republic}                                     
\centerline{$^{10}$Czech Technical University, Prague, Czech Republic}        
\centerline{$^{11}$Center for Particle Physics, Institute of Physics,         
                   Academy of Sciences of the Czech Republic,                 
                   Prague, Czech Republic}                                    
\centerline{$^{12}$Universidad San Francisco de Quito, Quito, Ecuador}        
\centerline{$^{13}$Laboratoire de Physique Corpusculaire, IN2P3-CNRS,         
                  Universit\'e Blaise Pascal, Clermont-Ferrand, France}       
\centerline{$^{14}$Laboratoire de Physique Subatomique et de Cosmologie,      
                  IN2P3-CNRS, Universite de Grenoble 1, Grenoble, France}     
\centerline{$^{15}$CPPM, IN2P3-CNRS, Universit\'e de la M\'editerran\'ee,     
                  Marseille, France}                                          
\centerline{$^{16}$IN2P3-CNRS, Laboratoire de l'Acc\'el\'erateur              
                  Lin\'eaire, Orsay, France}                                  
\centerline{$^{17}$LPNHE, IN2P3-CNRS, Universit\'es Paris VI and VII,         
                  Paris, France}                                              
\centerline{$^{18}$DAPNIA/Service de Physique des Particules, CEA, Saclay,    
                  France}                                                     
\centerline{$^{19}$IReS, IN2P3-CNRS, Universit\'e Louis Pasteur, Strasbourg,  
                France, and Universit\'e de Haute Alsace, Mulhouse, France}   
\centerline{$^{20}$Institut de Physique Nucl\'eaire de Lyon, IN2P3-CNRS,      
                   Universit\'e Claude Bernard, Villeurbanne, France}         
\centerline{$^{21}$III. Physikalisches Institut A, RWTH Aachen,               
                   Aachen, Germany}                                           
\centerline{$^{22}$Physikalisches Institut, Universit{\"a}t Bonn,             
                  Bonn, Germany}                                              
\centerline{$^{23}$Physikalisches Institut, Universit{\"a}t Freiburg,         
                  Freiburg, Germany}                                          
\centerline{$^{24}$Institut f{\"u}r Physik, Universit{\"a}t Mainz,            
                  Mainz, Germany}                                             
\centerline{$^{25}$Ludwig-Maximilians-Universit{\"a}t M{\"u}nchen,            
                   M{\"u}nchen, Germany}                                      
\centerline{$^{26}$Fachbereich Physik, University of Wuppertal,               
                   Wuppertal, Germany}                                        
\centerline{$^{27}$Panjab University, Chandigarh, India}                      
\centerline{$^{28}$Delhi University, Delhi, India}                            
\centerline{$^{29}$Tata Institute of Fundamental Research, Mumbai, India}     
\centerline{$^{30}$University College Dublin, Dublin, Ireland}                
\centerline{$^{31}$Korea Detector Laboratory, Korea University,               
                   Seoul, Korea}                                              
\centerline{$^{32}$CINVESTAV, Mexico City, Mexico}                            
\centerline{$^{33}$FOM-Institute NIKHEF and University of                     
                  Amsterdam/NIKHEF, Amsterdam, The Netherlands}               
\centerline{$^{34}$Radboud University Nijmegen/NIKHEF, Nijmegen, The          
                  Netherlands}                                                
\centerline{$^{35}$Joint Institute for Nuclear Research, Dubna, Russia}       
\centerline{$^{36}$Institute for Theoretical and Experimental Physics,        
                  Moscow, Russia}                                             
\centerline{$^{37}$Moscow State University, Moscow, Russia}                   
\centerline{$^{38}$Institute for High Energy Physics, Protvino, Russia}       
\centerline{$^{39}$Petersburg Nuclear Physics Institute,                      
                   St. Petersburg, Russia}                                    
\centerline{$^{40}$Lund University, Lund, Sweden, Royal Institute of          
                   Technology and Stockholm University, Stockholm,            
                   Sweden, and}                                               
\centerline{Uppsala University, Uppsala, Sweden}                              
\centerline{$^{41}$Lancaster University, Lancaster, United Kingdom}           
\centerline{$^{42}$Imperial College, London, United Kingdom}                  
\centerline{$^{43}$University of Manchester, Manchester, United Kingdom}      
\centerline{$^{44}$University of Arizona, Tucson, Arizona 85721, USA}         
\centerline{$^{45}$Lawrence Berkeley National Laboratory and University of    
                  California, Berkeley, California 94720, USA}                
\centerline{$^{46}$California State University, Fresno, California 93740, USA}
\centerline{$^{47}$University of California, Riverside, California 92521, USA}
\centerline{$^{48}$Florida State University, Tallahassee, Florida 32306, USA} 
\centerline{$^{49}$Fermi National Accelerator Laboratory, Batavia,            
                   Illinois 60510, USA}                                       
\centerline{$^{50}$University of Illinois at Chicago, Chicago,                
                   Illinois 60607, USA}                                       
\centerline{$^{51}$Northern Illinois University, DeKalb, Illinois 60115, USA} 
\centerline{$^{52}$Northwestern University, Evanston, Illinois 60208, USA}    
\centerline{$^{53}$Indiana University, Bloomington, Indiana 47405, USA}       
\centerline{$^{54}$University of Notre Dame, Notre Dame, Indiana 46556, USA}  
\centerline{$^{55}$Iowa State University, Ames, Iowa 50011, USA}              
\centerline{$^{56}$University of Kansas, Lawrence, Kansas 66045, USA}         
\centerline{$^{57}$Kansas State University, Manhattan, Kansas 66506, USA}     
\centerline{$^{58}$Louisiana Tech University, Ruston, Louisiana 71272, USA}   
\centerline{$^{59}$University of Maryland, College Park, Maryland 20742, USA} 
\centerline{$^{60}$Boston University, Boston, Massachusetts 02215, USA}       
\centerline{$^{61}$Northeastern University, Boston, Massachusetts 02115, USA} 
\centerline{$^{62}$University of Michigan, Ann Arbor, Michigan 48109, USA}    
\centerline{$^{63}$Michigan State University, East Lansing, Michigan 48824,   
                   USA}                                                       
\centerline{$^{64}$University of Mississippi, University, Mississippi 38677,  
                   USA}                                                       
\centerline{$^{65}$University of Nebraska, Lincoln, Nebraska 68588, USA}      
\centerline{$^{66}$Princeton University, Princeton, New Jersey 08544, USA}    
\centerline{$^{67}$Columbia University, New York, New York 10027, USA}        
\centerline{$^{68}$University of Rochester, Rochester, New York 14627, USA}   
\centerline{$^{69}$State University of New York, Stony Brook,                 
                   New York 11794, USA}                                       
\centerline{$^{70}$Brookhaven National Laboratory, Upton, New York 11973, USA}
\centerline{$^{71}$Langston University, Langston, Oklahoma 73050, USA}        
\centerline{$^{72}$University of Oklahoma, Norman, Oklahoma 73019, USA}       
\centerline{$^{73}$Brown University, Providence, Rhode Island 02912, USA}     
\centerline{$^{74}$University of Texas, Arlington, Texas 76019, USA}          
\centerline{$^{75}$Southern Methodist University, Dallas, Texas 75275, USA}   
\centerline{$^{76}$Rice University, Houston, Texas 77005, USA}                
\centerline{$^{77}$University of Virginia, Charlottesville, Virginia 22901,   
                   USA}                                                       
\centerline{$^{78}$University of Washington, Seattle, Washington 98195, USA}  
}                                                                             
%end                                                                          

\date{May 12, 2005}

% need to remove tex macros from abstract so it displays
% properly online
\begin{abstract}
  We present a measurement of the fraction $f_+$ of right-handed $W$
  bosons produced in top quark decays, based on a candidate sample of
  $t\bar{t}$ events in the lepton+jets decay mode. These data correspond to an
  integrated luminosity of $230$~pb$^{-1}$, collected by the D\O\
  detector at the Fermilab Tevatron $p\bar{p}$ Collider at
  $\sqrt{s}=1.96$~TeV.  We use a constrained fit to reconstruct the
  kinematics of the $t\bar{t}$ and decay products, which allows for the
  measurement of the leptonic decay angle $\theta^*$ for each event.  By
  comparing the $\cos\theta^*$ distribution from the data with those for
  the expected background and signal for various values of $f_+$, we
  find $f_+=0.00\pm0.13\mathrm{ (stat) }\pm0.07\mathrm{ (syst) }$.
  This measurement is consistent with the standard model prediction of
  $f_+=3.6\times10^{-4}$.
\end{abstract}


\maketitle

The top quark is by far the heaviest of the known fermions and is the
only one that has a Yukawa coupling of order unity to the Higgs boson
in the \SM.  
The top quark is also unique in that it decays through
the electroweak interaction before it can hadronize.
In the \SM, the top
quark decays via the $V-A$ charged current interaction, and almost
always to a $W$ boson and $b$ quark.  We search for evidence of new
physics in the \tWb decay by measuring the helicity of the $W$ boson.
The $W$ bosons produced from these decays are predominantly in either
a longitudinal or a left-handed helicity state with fractions \fzero
and \fminus, respectively.  For any linear combination of $V$ and $A$
currents at the $tWb$ vertex~\cite{fzero},
\begin{equation}
  \fzero\approx\frac{m_{t}^{2}}{2M_{W}^{2} +m_{t}^{2}+m_{b}^{2}}=0.703\pm0.012
\end{equation}
where $m_t$ is the mass of the top quark for which we use
$175\pm5~{\rm GeV}$ (consistent with the world average~\cite{pdg}),
$M_W$ is the mass of the $W$ boson, and $m_{b}$ is the mass of the
bottom quark.  In this analysis, we fix \fzero at $0.7$ and measure
the positive helicity (or right handed) fraction \fplus.  In the \SM, \fplus is
suppressed by a factor of $(m_b/m_t)^2$ and is predicted at
next-to-leading order to be $3.6\times10^{-4}$ ~\cite{fischer1}.  A
measurement of \fplus that differs significantly from this value would
be an unambiguous indication of new physics.  For example, an \fplus
value of $0.3$ would indicate a purely $V+A$ charged current
interaction.  A possible theoretical model that includes a $V+A$
contribution at the $tWb$ vertex is an $SU(2)_L\times SU(2)_R\times
U_Y(1)$ extension of the \SM~\cite{theory}.  Direct measurements of
the longitudinal fraction found
$\fzero=0.91\pm0.39$~\cite{helicityCDF2000} and
$\fzero=0.56\pm0.31$~\cite{helicityD0}.  A recent direct measurement
of \fplus set a limit of $\fplus<0.18$ at the $95\%$
\CL~\cite{helicityCDF2004}.  In addition, measurements of the $b
\rightarrow s\gamma$ decay rate have indirectly limited the $V+A$
contribution in top quark decays to less than a few
percent~\cite{sbg1}.  However, direct measurements of the $V+A$
contribution are still necessary because the limit from $b \rightarrow
s\gamma$ assumes that the electroweak penguin contribution is
dominant.

The angular distribution $\omega$ of the $W$ boson decay products with
weak isospin $I_3=-1/2$ (charged lepton or $d$, $s$ quark) in the rest
frame of the $W$ boson can be described by introducing the angle
\thetad with respect to the top quark direction~\cite{fzero}:
\begin{eqnarray}
  \omega(\cos\thetad) &=&\frac{3}{4}(1-\cos^2\thetad)\fzero+\frac{3}{8}(1-\cos\thetad)^2\fminus\nonumber\\
  &&+\frac{3}{8}(1+\cos\thetad)^2\fplus\;.
\end{eqnarray}
Due to backgrounds and reconstruction effects, the distribution of
$\costheta$ we observe differs from $\omega(\cos\thetad)$.  However,
the shape of the measured \costheta distribution depends on \fplus and
this dependence can be used to measure \fplus.  We do this by
selecting a data sample enriched in \ttbar events, reconstructing the
four vectors of the two top quarks and their decay products using a
kinematic fit, and then calculating \costheta.  This distribution in
\costheta is compared with templates for different \fplus values using
a binned maximum likelihood method.

The D\O\ detector~\cite{d0det} comprises three main systems: the
central-tracking system, the calorimeters, and the muon system.  The
central-tracking system is located within a 2~T solenoidal magnet.
The next layer of detection involves three liquid-argon/uranium
calorimeters: a central section (CC) covering
pseudorapidities~\cite{def_eta} $|\eta|\lesssim 1$, and two end
calorimeters (EC) extending coverage to $|\eta|\approx 4$, all housed
in separate cryostats.  The muon system is located beyond the
calorimetry, and consists of a layer of tracking detectors and
scintillation trigger counters before 1.8~T toroids, followed by two
more similar layers after the toroids.

This measurement uses a data sample recorded by the \dzero experiment
corresponding to $230\pm 15$~pb$^{-1}$ of \ppbar collisions at
$\sqrt{s}=1.96$~TeV.  We consider \ttbar candidate events selected in
the \ljets channel where one of the $W$ bosons from $t$ or $\bar{t}$
decays into an electron or muon and a corresponding neutrino and the
other $W$ boson decays hadronically.  The final state is therefore
characterized by one charged lepton ($e$ or $\mu$), at least four jets
(two of which are $b$ jets), and significant missing transverse energy
(\met).

Two separate analyses are performed and the results are combined.  One
analysis uses kinematic information to select \ttbar events
(``kinematic analysis'') and the other uses $b$ jet identification as
well as kinematic information in order to improve the signal to
background ratio (``$b$-tagged analysis'').  A $b$ jet is identified
by a displaced secondary vertex close to an associated
jet~\cite{xsec_btag}.  The kinematic analysis vetoes $b$-tagged events
to simplify the combination of results with the $b$-tagged analysis.
In both analyses, selected events arise predominantly from three
sources: \ttbar production, \wjets production, and multijet production
where one of the jets is misidentified as a lepton and spurious \met
appears due to mismeasurement of the transverse energy in the event.

The event selection~\cite{xsec_topo} requires an isolated lepton ($e$
or $\mu$) with transverse momentum $p_T>20$~$\mathrm{GeV}$, no other
lepton with $p_T>15$~$\mathrm{GeV}$ in the event, \met$ > 20$~GeV, and
at least four jets.  Leptons are categorized in two classes, ``loose''
and ``tight,'' the latter being a subset of the first.  Loose
electrons are required to have $|\eta|<1.1$ and are identified by
their energy deposition and isolation in the calorimeter, their
transverse and longitudinal shower shapes, and information from the
tracking system.  For tight identification, a discriminant combining
the above information must be consistent with the expectations for a
high-$p_T$ isolated electron.  Loose muons are identified using the
information from the muon and the tracking systems. They are required
to have $|\eta|<2.0$ and to be isolated from jets. Tight muons must
also pass stricter isolation requirements based on the energy of
calorimeter clusters and tracks around the muon. Only tight leptons
are used in the final event selection.  Jets are required to pass a
rapidity~\cite{def_eta} cut of $|y|<2.5$ and, in the kinematic
analysis, must have $p_T>20$~$\mathrm{GeV}$.  The requirement that a
$b$ jet is present significantly reduces the background contamination
in the $b$-tagged analysis and allows the use of a lower jet $p_T$ cut
of $p_T>15$~GeV which increases the efficiency for signal events.

The top quark and the $W$ boson four-momenta are reconstructed using a
kinematic fit which is subject to the following constraints: two jets
must form the invariant mass of the $W$ boson, the lepton and the \met
together with the neutrino $p_z$ component must form the invariant
mass of the $W$ boson, and the masses of the two reconstructed top
quarks must be equal to $175$~$\mathrm{GeV}$.  The $p_z$ component of the
neutrino is reconstructed by exploiting the fact that the masses of the two top
quarks are both set to be $175$~$\mathrm{GeV}$, and solving the resulting quadratic equation
for $p_z$~\cite{run1_topmass}. In the case where the two $p_z$
solutions lead to different results of the kinematic fit, the one with
the lower $\chi^2$ (of the fit) is kept.  Among the twelve possible
jet combinations, the solution with the minimal $\chi^2$ from the
kinematic fit is chosen; Monte Carlo studies show this yields the
correct solution in about $60\%$ of all cases.

The \ttbar signal events for seven different values of $f_+$,
$f_+=0.00, \ldots ,0.30$ in steps of $0.05$, are generated with the
\alpgen Monte Carlo (MC) program~\cite{alpgen} for the parton-level
process (leading order) and \pythia~\cite{pythia} for simulation of
subsequent hadronization. The mass of the top quark is set to
$m_t=175$~GeV.  As the interference term between $V-A$ and $V+A$ is
suppressed by the small mass of the $b$ quark and is therefore
negligible~\cite{bib_priv_koerner}, these samples can be used to
create \costheta templates for any $f_+$ value by a linear
interpolation of the templates. All seven templates from these samples
are normalized to unit area and a linear fit to the contents of each
\costheta bin as a function of \fplus is performed.  This procedure
effectively averages over statistical fluctuations in the generated MC
samples, thus providing a more precise model of the $\cos\thetad$
distribution.  The MC samples used to model events with $W$ bosons
produced in association with jets (\wjets) are also generated with
\alpgen, requiring the $W$ boson to decay leptonically. The
factorization scale $Q$ is set to $Q^2=M_W^2+\sum
m_T^2$~\cite{alpgen}.

To determine the number of multijet background events, we compare
samples selected with loose and tight leptons.  Going from loose to
tight samples decreases the number of events from $N_{\rm \ell}$ to
$N_t$.  The relative selection efficiency between the loose and the
tight lepton criteria is different for true leptons ($\epsilon_{\rm
  \ell}$) and jets faking an isolated lepton ($\epsilon_j$). We use
these efficiencies, known from data control samples~\cite{xsec_topo},
to estimate the number of multijet background events: $N_{\rm m} =
(\epsilon_{\rm \ell} N_{\rm \ell}-N_t)/ (\epsilon_{\rm
  \ell}-\epsilon_{j})$.  The kinematic analysis calculates $N_{\rm m}$
for each bin in the \costheta distribution from the data sample to
obtain the shape of the multijet \costheta templates.  For the
$b$-tagged analysis, the multijet template is formed from data events
after the event selection except that the leptons are required to
satisfy the loose and to fail the tight criteria.

To discriminate between \ttbar pair production and background, a
discriminant $\tld$ is built~\cite{xsec_topo} using input variables which exploit the
differences in event topology: $H_T$ (defined as the scalar sum of the
jet $p_T$ values), the minimum dijet mass of the jet pairs, the
$\chi^2$ from the kinematic fit, the centrality (defined as $H_T/H_E$
where $H_E$ is the sum of the jet energies)~\cite{run1_alljet}, $K_{\rm Tmin}^{\prime}$
(defined as the distance in $\eta-\phi$ space, where $\phi$ is the
azimuthal angle, between the closest pair of jets multiplied by the
$p_T$ of the lowest-$p_T$ jet in the pair and divided by the
transverse energy of the reconstructed $W$ boson)~\cite{run1_topmass}, and aplanarity and
sphericity (calculated from the four leading jets and the lepton). The
last two variables characterize the event shape and are defined, for
example, in Ref.~\cite{apla}.  Only the four leading jets in $p_T$ are
considered in computing these variables to reduce the dependence on
systematic effects from the modeling of soft radiation and underlying
event processes.  All of these variables are used for the discriminant
in the kinematic analysis.  Only $H_T$, centrality, the minimum dijet
mass, and $\chi^2$ are used in the $b$-tagged analysis.  The
discriminant is built separately for the kinematic and $b$-tagged
analyses, using the method described in
Refs.~\cite{xsec_topo,run1_topmass}.  The distributions of signal
($S$) and background ($B$) events in each of the above variables are
normalized to unity. For each variable $v_i$ we fit a polynomial to
the logarithm of $S/B$ as a function of $v_i$.  The discriminant is
defined as:
\begin{equation}
  \tld(v_1,v_2,\ldots) =
  \frac{\exp\left(\sum_i
      {[\ln
        (S(v_i)/B(v_i))
        ]_{\rm fit}
      }
    \right)
  }{\exp\left(\sum_i
      {[\ln
        (S(v_i)/B(v_i))
        ]_{\rm fit}
      }
    \right) + 1
  }\;.
\end{equation}
We select events for which $\tld>0.6$ in the kinematic analysis, and
$\tld>0.25$ in the $b$-tagged analysis.  These values are chosen to
minimize the expected statistical uncertainty in the measurement of
\fplus as determined by simulations of the analysis.

\begin{figure}
  \includegraphics[trim=0 15 0 7,scale=0.7]{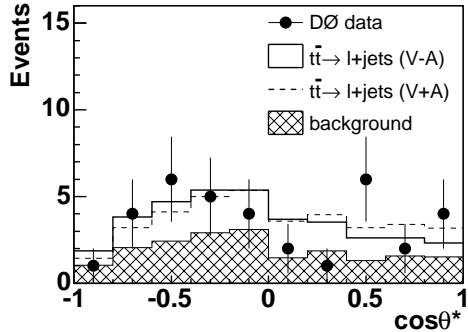}
  \caption{\label{fig_cost_topo}\costheta distribution observed in the
    kinematic analysis. The \SM ~prediction is shown as the solid line,
    while a model with a pure $V+A$ interaction would result in the
    distribution given by the dashed line.}
\end{figure}

\begin{figure}
  \includegraphics[trim=0 15 0 7,scale=0.7]{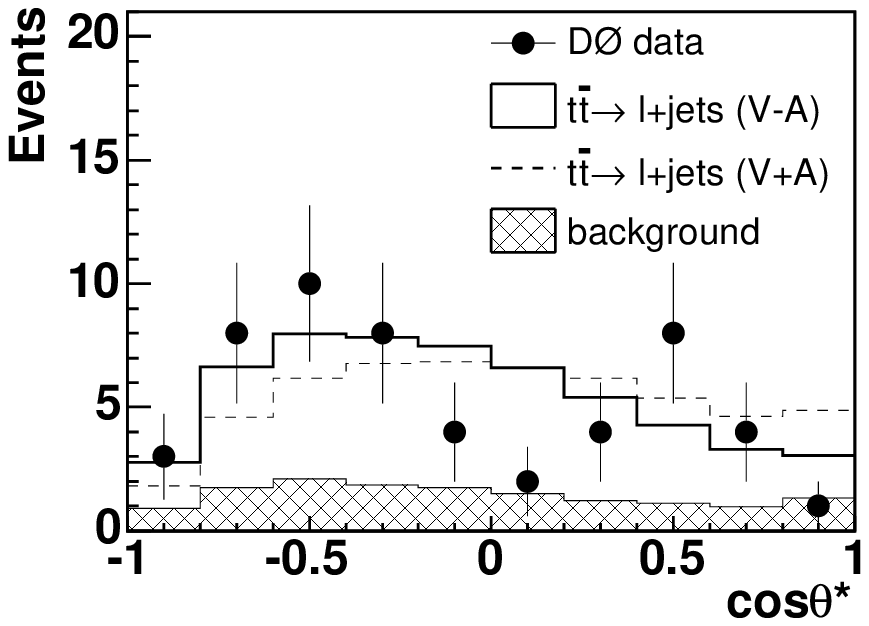}
  \caption{\label{fig_cost_btag}\costheta distribution observed in the
    $b$-tagged analysis. The \SM ~prediction is shown as the solid
    line, while a model with a pure $V+A$ interaction would result in
    the distribution given by the dashed line.}
\end{figure}

We then perform a binned maximum likelihood fit to compare the
observed \tld ~distribution in the data to the sum of the distributions
expected from \ttbar, \wjets, and multijet events.  The number of
multijet events is constrained to a Poisson distribution with mean
$N_{\rm m}$.  The likelihood is then maximized with respect to the
number of \ttbar, \wjets, and multijet events.  We multiply these
numbers by the efficiency for each type of event to pass the \tld
selection to determine the composition of the sample used for
measuring \costheta.  Table~\ref{tab_sig_bkg_numbers} lists the
composition of each sample as well as the number of observed events in
the data.  The \costheta distribution obtained in data after the full
selection is shown in Fig.~\ref{fig_cost_topo} for the kinematic and
in Fig.~\ref{fig_cost_btag} for the $b$-tagged analysis.

\begin{table}
  \caption{\label{tab_sig_bkg_numbers}%
Number of events observed for each component (signal and backgrounds) 
of the kinematic and $b$-tagged samples, and the number of data events, 
after the cut on the discriminant \tld discussed in the text.  The fits are 
made before the final cut on \tld, so the sum of the components 
need not agree exactly with the observed numbers of events.}
\begin{ruledtabular}
  \begin{tabular}{lcc}
    Event Class & Kinematic & $b$-tagged \\
    \hline
    \ttbar    & $16.5 \pm 5.8$ & $40.8 \pm 8.1$ \\
    \wjets      & $14.3 \pm 3.0$ & $11.5 \pm 4.1$ \\
    Multijet & $\phantom{1} 5.0 \pm 2.1$ & $\phantom{8} 1.5 \pm 0.5$ \\ 
    Data & 35 & 52 \\
  \end{tabular}
\end{ruledtabular}
\end{table}

A binned maximum likelihood fit of signal and background \costheta
templates to the data was used to measure \fplus.  We compute the
binned Poisson likelihood ($L(\fplus)$) of the data to be consistent
with the sum of signal and background templates, normalized to the
numbers given in Table~\ref{tab_sig_bkg_numbers}, at each of the seven
chosen \fplus values.  In both analyses, a parabola is fit to the
$-\ln[L(\fplus)]$ points to determine the likelihood as a function of
$\fplus$.

\begin{figure*}
  \begin{tabular}{ccc}
    \includegraphics[trim = 2 15 16 0,scale=0.65]{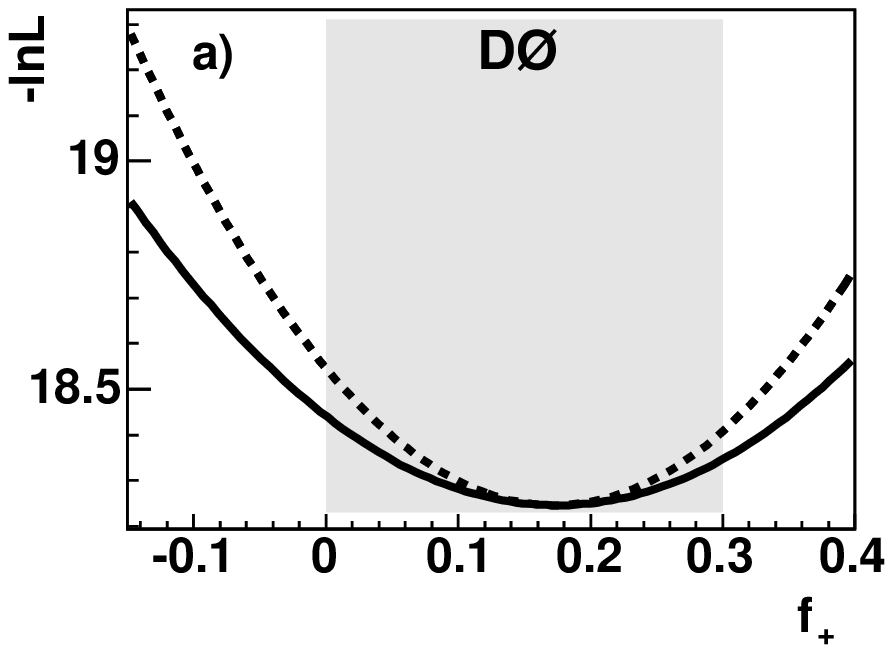} & 
    \includegraphics[trim = 2 15 16 0,scale=0.65]{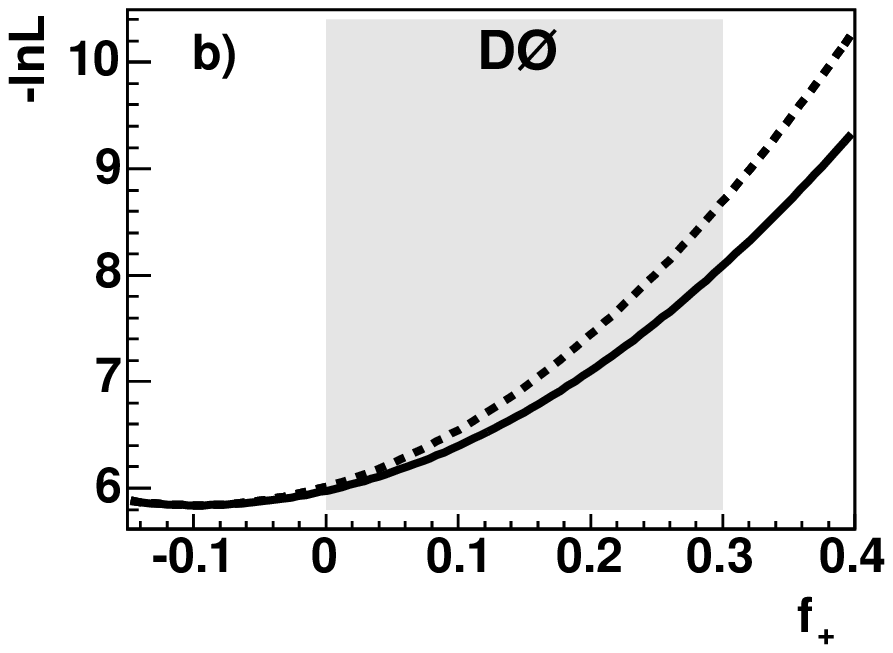} &
    \includegraphics[trim = 2 15 16 0,scale=0.65]{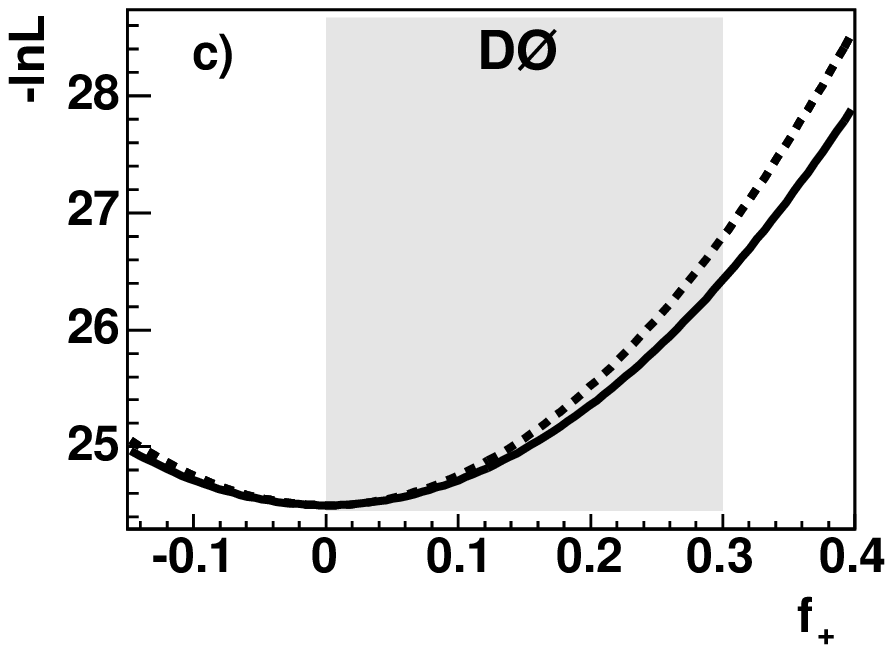} 
  \end{tabular}
  \caption{\label{fig_result} $-\ln L$ curve obtained in the a)
    kinematic analysis, b) $b$-tagged analysis, and c) kinematic and
    $b$-tagged analyses combined. The dashed line includes only the
    statistical uncertainty while the solid line also includes the
    systematic uncertainties. The physically allowed region for \fplus
    is indicated by the grey area.}
\end{figure*}

Systematic uncertainties are evaluated in ensemble tests by varying
the parameters (see Table~\ref{tab_systematic}) which can affect the
shape of the \costheta distributions or the relative contribution from
the three sources (\ttbar, \wjets and QCD).  Ensembles are formed by
drawing events from a model with the parameter under study varied.
These are compared to the standard \costheta templates in a maximum
likelihood fit.  The average shift in the resulting \fplus value is
taken as the systematic uncertainty and is shown in
Table~\ref{tab_systematic}.  The total systematic uncertainty is then
taken into account in the likelihood by convoluting the latter with a
Gaussian with a width that corresponds to the total systematic
uncertainty.

The top quark mass and the jet energy calibration (JEC) are the
leading sources of systematic uncertainty.  The mass of the top quark
has been varied by $\pm5$~$\mathrm{GeV}$ with respect to
$m_t=175$~$\mathrm{GeV}$ and the JEC by $\pm1\sigma$ around the
nominal value.  The statistical uncertainty on the \costheta templates
has been taken as a systematic uncertainty.  It is estimated by
fluctuating them according to their statistical uncertainty.
Uncertainties in the modeling of the $b$-tag algorithm lead to
uncertainties in the flavor composition of the $W+$jets background and
in the \costheta distribution itself due to the $p_T$ and $\eta$
dependence of the $b$-tag algorithm~\cite{xsec_btag}.
An uncertainty in the flavor composition translates into a 
different shape of the \costheta distribution and a difference in the
signal to background ratio.
  In order to
estimate the systematic uncertainty due to gluon radiation in \ttbar
events, an alternative signal sample of $t\bar{t}$+jet has been
generated with \alpgen, and mixed with the default \ttbar sample using
the leading order cross sections for both processes. Effects of the
choice of factorization scale $Q$ in the generation of the $W$+jets
events have been evaluated by using a sample where $Q^2=\langle
p_T\rangle^2$~\cite{alpgen}.  There is a systematic uncertainty due to
the final sample composition obtained by the fit to the discriminant
\tld.  The kinematic analysis treats this uncertainty as a statistical
uncertainty and includes it in the definition of the likelihood as
described in Ref.~\cite{ext_lh} while in the $b$-tagged analysis this
uncertainty is studied by changing the compositions within their
errors.  The difference found between the input \fplus value and the
reconstructed \fplus value in ensemble tests is taken as systematic
uncertainty on the calibration of the analysis.
\begin{table}
  \caption{\label{tab_systematic}Systematic uncertainties on \fplus for the 
    two independent analyses and for the combination.}
  \begin{ruledtabular}
    \begin{tabular}{lccc}
      Source & Kinematic & $b$-tagged & Combined\\
      \hline
      Jet energy calibration & 0.03 & 0.04 & 0.04\\
      Top quark mass         & 0.04 & 0.04 & 0.04\\
      Template statistics    & 0.05 & 0.02 & 0.03\\ 
      $b$-tag                & 0.03 & 0.02 & 0.02\\
      \ttbar model           & 0.01 & 0.02 & 0.02\\
      $W+$jets model         & 0.01 & 0.01 & 0.01\\
      Sample composition     & ---  & 0.02 & 0.01\\
      Calibration            & 0.01 & 0.01 & 0.01\\
      \hline
      Total                  & 0.08 & 0.07 & 0.07
    \end{tabular}
  \end{ruledtabular}
\end{table}

The result of the maximum likelihood fit to the \costheta distribution
observed in the data is shown in Figs.~\ref{fig_result}(a) and~(b) for
the kinematic and $b$-tagged samples, respectively.  The statistical
uncertainties from the two individual analyses are $0.22$ for the
kinematic and $0.17$ for the $b$-tagged analysis.  The
$-\ln[L(\fplus)]$ curves for the kinematic and $b$-tagged measurements
are combined, as shown in Fig.~\ref{fig_result}(c).  The systematic
uncertainties are assumed to be fully correlated except for the
systematics on calibration of the individual analyses which are
uncorrelated, and the Monte Carlo model systematics which are
partially correlated.  Assuming a fixed value of $0.7$ for \fzero, the
combined result for \fplus is:
\begin{equation}
  \fplus=0.00\pm0.13\mathrm{ (stat) }\pm0.07\mathrm{ (syst) }.
\end{equation}
The observed combined statistical uncertainty ($0.13$) is in good
agreement with the expectation ($0.12$) inferred from ensemble tests.
We also calculate a Bayesian confidence interval (using a flat prior
distribution which is non-zero only in the physically allowed region
of $\fplus=0.0-0.3$) which yields
\begin{equation}
  \fplus<0.25\mathrm{\ @\ }95\%\mathrm{\ \CL}
\end{equation}

The $W$ boson positive helicity fraction \fplus that we have measured
in \ttbar decays in the \ljets ~channel is consistent with the standard
model prediction of $\fplus=3.6\times10^{-4}$~\cite{fischer1}.

% acknowledgement_paragraph_r2.tex 5/10/05
%
We thank the staffs at Fermilab and collaborating institutions, and
acknowledge support from the DOE and NSF (USA); CEA and CNRS/IN2P3
(France); FASI, Rosatom and RFBR (Russia); CAPES, CNPq, FAPERJ, FAPESP
and FUNDUNESP (Brazil); DAE and DST (India); Colciencias (Colombia);
CONACyT (Mexico); KRF (Korea); CONICET and UBACyT (Argentina); FOM
(The Netherlands); PPARC (United Kingdom); MSMT (Czech Republic); CRC
Program, CFI, NSERC and WestGrid Project (Canada); BMBF and DFG
(Germany); SFI (Ireland); Research Corporation, Alexander von Humboldt
Foundation, and the Marie Curie Fellowships.


\begin{thebibliography}{99}
\bibitem{def_eta} Rapidity $y$ and pseudorapidity $\eta$ are defined
  as functions of the polar angle $\theta$ with respect to the proton
  beam and the parameter $\beta$ as $y(\theta,\beta)\equiv
  \frac{1}{2}\ln[(1+\beta\cos\theta)/(1-\beta\cos\theta)]$ and
  $\eta(\theta)\equiv y(\theta,1)$, where $\beta$ is the ratio of a
  particle's momentum to its energy.
\end{thebibliography}
\end{document}